\pdfoutput=1
\documentclass[aps, pra,reprint,longbibliography,superscriptaddress]{revtex4-1}
\usepackage{amsmath}
\usepackage{bm}
\usepackage{amssymb}
\usepackage{graphicx}
\usepackage{times}
\usepackage{hyperref}
\usepackage{verbatim}
\usepackage{color}
\usepackage{cleveref}
\usepackage[normalem]{ulem}

\allowdisplaybreaks[3]

\newcommand{\latname}{$8^2.10$-$a$}
\begin{document}

\title{Entanglement in 3D Kitaev Spin Liquids}

\author{S. Matern}
\affiliation{SUPA, School of Physics and Astronomy, University of St Andrews, North Haugh, St Andrews, Fife KY16 9SS, United Kingdom}
\affiliation{Institute for Theoretical Physics, University of Cologne, 50937 Cologne, Germany}
\author{M. Hermanns}
\affiliation{Department of Physics, University of Gothenburg, SE 412 96 Gothenburg, Sweden}
\affiliation{Institute for Theoretical Physics, University of Cologne, 50937 Cologne, Germany}

\date{\today}

\begin{abstract}
Quantum spin liquids  are highly fascinating quantum liquids in which the spin degrees of freedom fractionalize. 
An interesting class of spin liquids are the exactly solvable, three-dimensional Kitaev spin liquids. 
Their fractionalized excitations are Majonara fermions, which may exhibit a variety of topological band structures --- ranging from topologically protected Weyl semi-metals over nodal semi-metals to systems with Majorana Fermi surfaces. 
We study the entanglement spectrum of such Kitaev spin liquids and verify that it is closely related to the topologically protected edge spectrum. 
Moreover, we find that in some cases the entanglement spectrum  contains even more information about the topological features than the surface spectrum, and thus provides a simple and reliable tool to probe the topology of a system. 
\end{abstract}

\pacs{\ }
\maketitle

%\noindent 
\section{Introduction}
The way we understand and describe phases of matter in condensed matter systems has changed substantially during the last few decades. 
Ever since the discovery of the quantum Hall effect --- both integer and fractional~\cite{klitzingdiscovery,tsuidiscovery} --- it has become apparent that phases of matter cannot be uniquely characterized by (local) order parameters.
Instead, concepts from other fields --- such as topology from mathematics and entanglement from quantum information --- have proven useful. 
For instance, it was shown that the accurate quantization of the  Hall conductance in the integer quantum Hall effect is linked to a \emph{topological invariant}, the Chern number, of the free fermion bulk band structure~\cite{TKNN}. 
Such a feature is not unique to the quantum Hall effect: ever since the seminal paper of Kane and Mele introducing topological insulators~\cite{Kane2005quantum}, (noninteracting) \emph{topological phases of matter} have been studied extensively and are by now very well understood~\cite{Altland1997classification,Schnyder2008classification,Kitaev2009periodic,Chiu2016classification}.
Furthermore, even gapless systems can be topological in the sense that the topological invariant protects both the gapless bulk nodes and the corresponding gapless surface states~\cite{Chiu2016classification}; prominent examples are nodal line semimetals~\cite{Burkov2011topological} and Weyl semimetals~\cite{Wan2011topological}.

Even more interesting are \emph{interacting} topological phases, because they may harbor highly exotic quasiparticle excitations~\cite{leinaas1977theory,Wilczek1982quantum}. 
Such phases are often called topologically ordered~\cite{Wen1995topological} or long-range entangled~\cite{QImeetsCondMat}, which implies that their \emph{topological entanglement entropy} is non-vanishing~\cite{Hamma1,LevinWen,KitaevPreskill,Haque2007entanglement}.
The latter is currently one of the most powerful tools to verify in numerical simulations whether a ground state is topologically ordered or not. 
The entanglement entropy is obtained by partitioning the system into two spatial parts --- called $A$ and $B$ in the following --- and computing the von Neumann entropy of part $A$,
\begin{align}\label{eq:vNeumann}
S_A&=-\mbox{Tr}\left( \rho_A\ln\rho_A\right)\nonumber\\
&=\alpha \ell-\gamma,
\end{align}
where $\rho_A$ is the reduced density matrix of $A$.  
For gapped phases, the entanglement entropy obeys an area law, i.e. it grows as the area/length of the boundary $\ell$ between $A$ and $B$. The coefficient of the area law (here denoted by $\alpha$) is a nonuniversal constant and depends on microscopic details.  
For topologically ordered systems, the entanglement entropy has an additional, sub-leading constant contribution $\gamma$ that is independent of microscopic details. 
It can be identified with the logarithm of the total quantum dimension~\cite{LevinWen,KitaevPreskill}, which gives valuable insight on which topological field theory captures the low-energy, long-wavelength physics of the system. 
However, as it only measures a single quantum number, it is not sufficient to determine the topological order uniquely. 
A potentially more powerful diagnostic tool is the \emph{entanglement spectrum} (ES)~\cite{Li2008entanglement}, which is defined as the spectrum of the entanglement Hamiltonian $H_\text{ent}=-\ln(\rho_A)$.
While it is a property solely of the ground state, it contains information about the excitations of the systems --- both edge and bulk~\cite{Sterdyniak2011extracting,chandran2011bulk,Qi2012general}, depending on how the Hilbert space is bipartitioned.
By now, the ES has proven to be a useful tool for a variety of systems~\cite{Pollmann2010entanglement,Regnault2011fractional,Lauchli2013unpublished,He2014chiral}.

A prominent example where the entanglement entropy fails to distinguish two topologically ordered states from each other is the Kitaev honeycomb model~\cite{Kitaev2006anyons} --- a paradigmatic model of a quantum spin liquid.
This model considers spin-1/2 local moments on the sites of the honeycomb lattice with a highly anisotropic Ising-like interaction
\begin{align} 
H=-\sum_{\langle j,k\rangle }J_{\gamma}\sigma_j^\gamma\sigma_k^\gamma,
\label{eq:kitaev}
\end{align}
where the sum is over all nearest neighbors $j,k$ and the  local easy-axis $\gamma=x,y,z$ depends on the  direction of the nearest-neighbor bond $\langle j,k\rangle$. 
In this exactly solvable model, the spins fractionalize into itinerant Majorana fermions and a static $\mathbb{Z}_2$ gauge field. 
The latter are generically gapped, while the former may be gapped or gapless depending on the parameters. 
The Kitaev honeycomb model has two, distinct gapped phases: (i) a toric-code like phase, when one of the coupling constants is larger than the sum of the other two and (ii) a nonabelian phase with Ising anyon excitations, when all the coupling constants are approximately the same ($J_x\approx J_y \approx J_z$) and time-reversal symmetry is broken (e.g. by applying an external magnetic field)~\cite{Kitaev2006anyons}.   
While these two gapped phases have very different properties, their topological entanglement entropy happens to be identical. 
Yao and Qi gave a physical explanation of this feature by noting that the entanglement entropy of this model can be written as a sum of two terms ---  one capturing the effect of the Majorana fermions and the other the effect of the gauge field~\cite{Yao2010entanglement}. 
Only the latter contributes to the topological entanglement entropy, while it is the former that actually determines which type of Kitaev spin liquid (KSL) is realized in the system.
They also showed that the  ES can distinguish the two phases. Moreover, it can be efficiently computed using methods for noninteracting systems~\cite{Peschel2003calculation}. 

Entanglement spectra have mainly been studied for gapped systems, even though they also have interesting properties in gapless systems~\cite{Hermanns2014entanglement}. 
Here, we analyse entanglement spectra for gapless three-dimensional KSLs and show that topologically protected zero-energy surface modes give rise to  modes in the ES that reside  at `entanglement energy' 1/2.
Entanglement spectra are most often easier to interpret and analyze than the corresponding surface spectra, because the latter may have accidental modes near zero-energy, while the former do not have accidental modes at 1/2. 
Moreover, entanglement spectra can in some cases even distinguish whether non-topological features have a topological origin.
Our results are not only valid for KSLs, but can easily be generalized to noninteracting gapless fermionic systems

This manuscript is organized as follows. First, we give a short review of KSLs in section~\ref{sec:KSL} and entanglement spectra of noninteracting systems in section~\ref{sec:ESnoninteracting}. We then proceed to discuss the features of (gapless) topological band structures in the ES, using some KSLs as explicit examples in section \ref{sec:ESvsSurface}. In section \ref{sec:topfeatures} we argue that the ES can reveal interesting weak topological features of the system that cannot be identified by the surface spectrum.  Details on the various examples of KSLs can be found in the appendices.  
%-------------------

%-------------------
\section{Kitaev spin liquids}\label{sec:KSL}

\begin{figure}[t]
\centering
\includegraphics{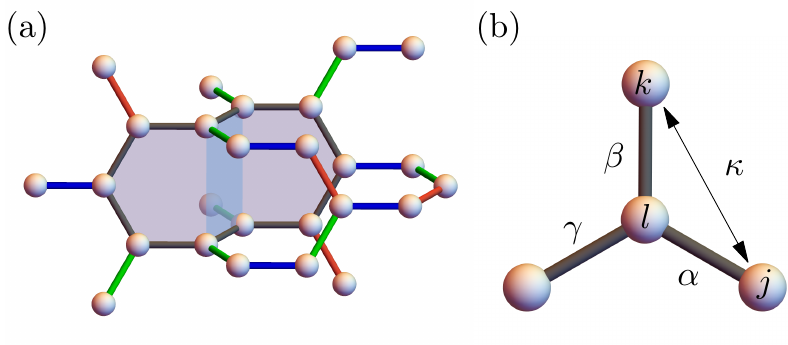}
\caption{(a) Example of a closed loop of length 10 (depicted in gray) in the hyperhoneycomb lattice. (b) The (time-reversal breaking) three-spin-interaction of Eq.~\eqref{eq:mag}, where the arrow indicates the induced next-nearest-neighbor hopping of the Majorana fermions. }
\label{fig:loop}
\end{figure}

We start by giving a short review  on the solution of the Kitaev honeycomb model~\cite{Kitaev2006anyons} in Eq.~\eqref{eq:kitaev}, which can be generalized to any tricoordinated lattice. 
For further details, we refer the reader to one of  the various reviews on the subject, see for instance Ref.s~\cite{Kitaev2009topological,Nussinov2015compass,Hermanns2017physics}.
We can represent the spin-1/2 operators by introducing four Majorana fermions per site ($b_j^x$, $b_j^y$, $b_j^z$, and $c_j$):
\begin{align}
\sigma_j^\gamma&=ib_j^\gamma c_j.
\label{eq:sigmaMaj}
\end{align}
Note that the local Hilbert space of four Majorana fermions is four-dimensional (per site), whereas the local Hilbert space of the spins is only two-dimensional. 
By enlarging the Hilbert space, we have introduced unphysical degrees of freedom that can be identified with \emph{emergent} $\mathbb{Z}_2$ gauge fields, $\hat{u}_{jk}=ib_j^\gamma b_k^\gamma$, living on the $\gamma$-type bonds $\langle j,k\rangle$. 
Physical states are required to fulfill
\begin{align}
D_j |\mbox{phys}\rangle&\equiv b_j^x b_j^y b_j^z c_j |\mbox{phys}\rangle =+1 |\mbox{phys}\rangle,
\label{eq:D}
\end{align}
for all $j$, which ensures that Eq.~\eqref{eq:sigmaMaj} reproduces the spin algebra in the physical Hilbert space. 
Applying $D$ on a state can be thought of as a gauge transformation of the  $\mathbb{Z}_2$ gauge field, and the condition in Eq.~\eqref{eq:D} simply means that physical states are \emph{gauge invariant}. 
Note that the eigenvalues of the bond operators are not physical. However, we can define physical quantities by considering products of bond operators along a closed loop $\ell$ in the lattice [see Fig.~\ref{fig:loop}(a)],  
\begin{align}
W_\ell&=\prod_{\langle j,k\rangle \in \ell} (-i \hat u_{jk}). 
\end{align}
The loop operator $W_{\ell}$ has eigenvalues $\pm 1$ and measures the flux through the loop. In the following, the eigenvalue $+1$ is referred to as zero flux and $-1$ as $\pi$ flux. 

Rewriting the spin Hamiltonian \eqref{eq:kitaev} in terms of the Majorana fermions yields 
\begin{align}
H_{MF}&=i \sum_{\langle j,k\rangle}J_\gamma \hat{u}_{jk} c_j c_k. 
\label{eq:bareKitaev}
\end{align}
Note that the bond operators $\hat{u}_{jk}$ commute among themselves as well as with the Hamiltonian, so we can simply fix their eigenvalues, {\it i.e.} fix the gauge. 
In order to find the ground state, we need to identify the $\mathbb{Z}_2$ flux sector that minimizes the energy. 
For a few lattices, this can be done rigorously using Lieb's theorem \cite{Lieb1994flux}, but most often one needs to resolve to numerical calculations. 
For the lattices discussed in this manuscript, the ground state flux configurations are such that  loops of length 2 mod 4 have zero flux and loops of length 0 mod 4 have $\pi$ flux \cite{Obrien2016classification,Yamada2017crystalline}. 
Once the flux sector is fixed,  Eq.~\eqref{eq:bareKitaev} reduces to a noninteracting Hamiltonian, where Majorana fermions hop in a static, gapped $\mathbb{Z}_2$ gauge field.
This Majorana Hamiltonian can be gapped or gapless, and its band structure determines the nature of the KSL. 
In the remainder of the manuscript, we focus on this Hamiltonian.

We are not only interested in the properties of the bare Kitaev model, but also in the effect of breaking time-reversal symmetry or lattice symmetries. 
The simplest time-reversal breaking term is given by the interaction of three neighboring spins
\begin{align}
H_{\text{mag}}\sim-\kappa\sum_{jkl}\sigma_j^\alpha\sigma_k^\beta\sigma_l^\gamma, 
\label{eq:mag}
\end{align}
as it does not destroy the exact solvability of the model. 
Here, $\alpha$, $\beta$ and $\gamma$ denote the types of the bonds surrounding the central site $l$, as shown in Fig.~ \ref{fig:loop}(b).
This three-spin interaction has a natural interpretation as the projection of a Zeeman term, $\sum_{j} \mathbf{h}\cdot \pmb{\sigma}_j$, into the ground state flux sector~\cite{Kitaev2006anyons}. 
In this case, the coupling constant is given by $\kappa \sim h_x h_y h_z/\Delta_E^2$, where $\Delta_E$ denotes the  gap of the $\mathbb{Z}_2$ gauge field. 
In terms of the Majorana fermions, the three-spin interaction of Eq.~\eqref{eq:mag} is nothing but a next-nearest neighbor hopping term.
\begin{align}
H_{\text{mag}}-i  \kappa \sum_{\langle \! \langle j,k \rangle \! \rangle} \tilde u_{jk} c_j c_k
\end{align}
with $\tilde u_{jk}=-\epsilon^{\alpha\beta\gamma}$ and $\langle \! \langle j,k \rangle \! \rangle$ denoting the bond between next-nearest neighbors $j$ and $k$. 

Depending on the symmetries and the underlying lattice, the Kitaev model can show a host of different quantum spin liquid phases \cite{Obrien2016classification}. 
In this manuscript, we focus on a few lattices that showcase the different gapless spin liquid phases and their corresponding entanglement features. 
For more details on the various KSLs, we refer the reader to the appendices and Refs.~\cite{Obrien2016classification,Yamada2017crystalline}.

\section{Entanglement spectra for noninteracting systems} 
\label{sec:ESnoninteracting}
The most common method to identify topologically ordered phases, such as quantum spin liquids, is the \emph{ entanglement entropy} \cite{LevinWen,KitaevPreskill}. 
Its subleading contribution --- the topological entanglement entropy ---  measures the total quantum dimension of the underlying effective topological field theory description. 
For trivial gapped systems --- in particular for noninteracting ones --- the topological entanglement entropy vanishes. 

For the Kitaev honeycomb model  the entanglement entropy  can be written as the sum of two terms \cite{Yao2010entanglement}
\begin{align}
S&=S_G +S_{MF},
\end{align}
where $S_G$ depends only on the gauge field and contributes $\log(2)$ to the topological entanglement entropy as is appropriate for a $\mathbb{Z}_2$ gauge theory. 
The second term contains the contribution of the free Majorana fermions and can, therefore, only contribute  to the nonuniversal coefficient of the area law, but not to the topological entanglement entropy. 
This behavior is generic for KSLs; in particular, it does not depend on the details of the underlying lattice or on its dimensionality \cite{Mondragon2014entanglement}. 
Consequently,  the topological entanglement entropy is the same for all KSLs (with the possible exception of those where the $\mathbb{Z}_2$ flux model is frustrated \cite{Obrien2016classification}). 
The ES, on the other hand, is sensitive to the topological invariants of the effective Majorana Hamiltonian~\cite{Fidkowski2010entanglement}, and  can, thus, provide valuable information for the various KSLs.

\begin{figure}
	\centering
	\includegraphics{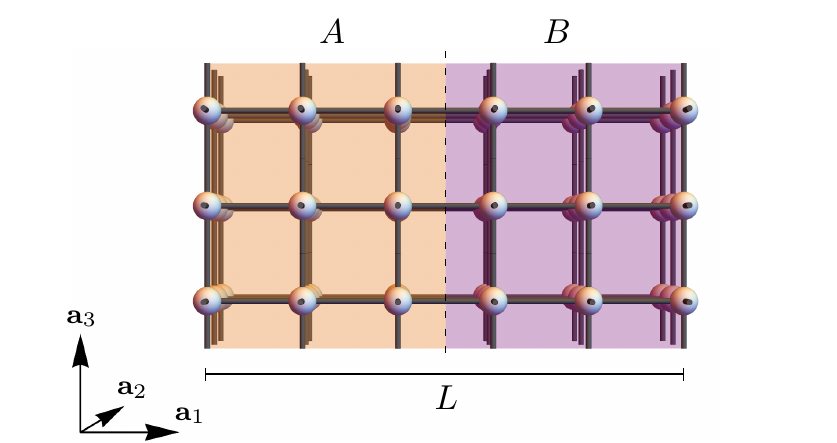}
	\caption{
		Schematic visualization of the spatial bipartition used in the remainder of the manuscript. Due to the periodic boundary conditions along the $\mathbf a_1$ direction, there are two virtual surfaces: one at $r_1=0$ and the other at $r_1=L/2$. }
	\label{fig:bipartition}
\end{figure}

In the following, we use the method by Peschel \cite{Peschel2003calculation} to compute the ES for the Majorana fermions. 
For a noninteracting tight-binding Hamiltonian, the reduced density matrix is uniquely determined by the one-particle correlation function, 
\begin{align}
	C^{\alpha\beta}_{nm}&=\langle 0|c^\dagger_{\alpha,n} c_{\beta,m} |0 \rangle, 
\end{align}
where $\alpha$ and $\beta$ denote the set of internal quantum numbers per site (such as spin, orbital, etc.), $n$ and $m$ denote sites that are restricted to lie in subsystem $A$, and $|0\rangle$ denotes the ground state. 
We always choose a bipartition such that two directions remain periodic, e.g. $\mathbf  a_2$ and $\mathbf a_3$ in the  schematic visualization in Fig.~\ref{fig:bipartition}, so that we can use their corresponding momenta $\tilde{\mathbf k}=(k_2,k_3)$ to label the eigenvalues. 

After a partial Fourier transform, the correlation matrix becomes 
\begin{align}
C^{\alpha\beta}_{nm}(\tilde{\mathbf k})&=\left\langle c^\dagger_\alpha (n,\tilde{\mathbf k})c_\beta (m,\tilde{\mathbf k})\right\rangle\notag\\
&=\sum_{j,E_j<0}\sum_{q} \text{e}^{i q(m-n)}U^\dagger_{\beta j}(\mathbf k)U_{j\alpha}(\mathbf k),
\end{align}
where $\alpha$ and $\beta$ denote the sites within the unit cell and the sum over $j$ runs over all the occupied states ($E_j<0$). The unitary matrix $U$ is defined by the eigenstates
$\eta_\alpha(\mathbf{k})=\sum_{\beta}U_{\alpha \beta} c_{\beta}(\mathbf{k})$, with $\mathbf k=(q,\tilde{\mathbf k})$. 
In the following, we refer to the  eigenvalues $\zeta$ of the matrix $C^{\alpha\beta}_{nm}$ as the ES.

The full ES of the subsystem $A$ consists of $M\cdot L_A$ bands, where $L_A$ is length of region $A$, and $M$ is the number of sites in the unit cell. 
For the calculation of the correlation matrix, the length of the subsystem $L_A$ was chosen as $L/2$ throughout the paper, with $L^3$ being the system size. 
Most of the entanglement bands are located around 0 (1) corresponding to an empty (occupied) state. 
The relevant bands we want to study have values $0<\zeta<1$, because they turn out to be related to the topologically protected surface modes. 
For noninteracting topological insulators and superconductors, the ES is, in fact, identical to the spectrally flattened surface spectrum, as shown by Fidkowski~\cite{Fidkowski2010entanglement}. 
For each topological surface band that connects valence and conduction bands, there is a corresponding band in the ES connecting the bands at entanglement energy 0 and 1.
Consequently, ES modes at energy $\zeta=1/2$ are in one-to-one correspondence to the surface modes at zero energy. 

\begin{figure*}[t]
	\centering
	\includegraphics{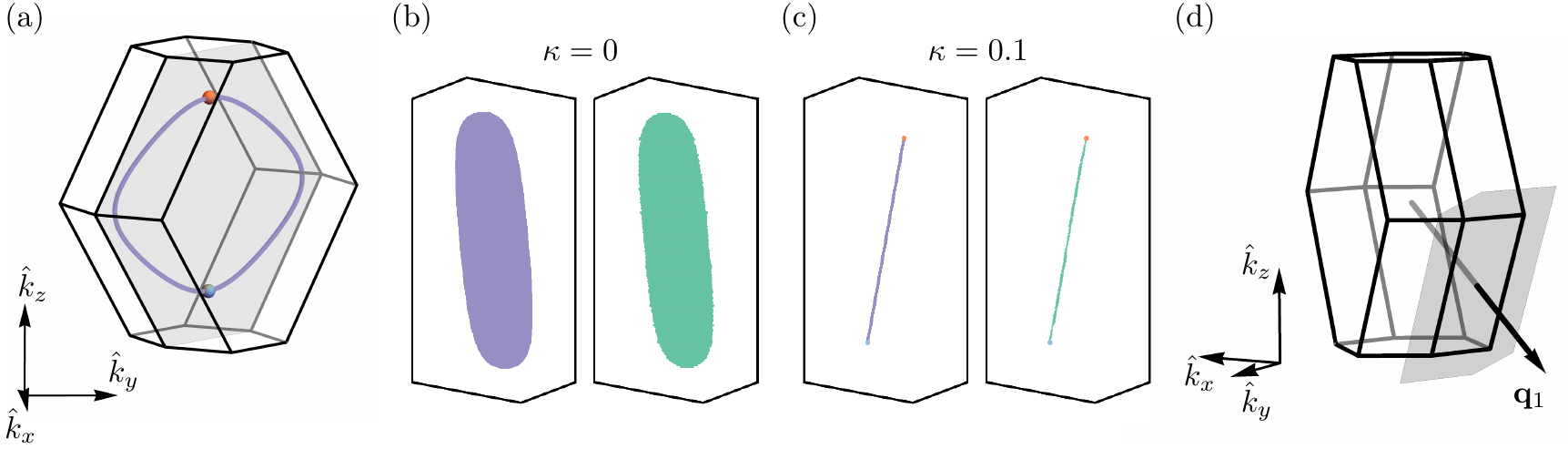}
	\caption{
		(a) Full BZ of the (10,3)b lattice. The time-reversal symmetric system harbors a nodal line (purple). Breaking time-reversal gaps the line into two WPs, marked by the orange and blue point. 
		The middle two panels show the zero-energy states of the surface spectrum (purple) and the $\zeta=1/2$ states of the ES (green) for (b) the time-reversal invariant system and (c) the time-reversal broken system.
		The surface BZ is shown in (d) for open boundary conditions along $\mathbf{a}_1$.}
	\label{fig:hhoney}
\end{figure*}

Before proceeding to describe the ES for the various KSLs, let us note that for all entanglement spectra we used the ground state in the extended Hilbert space without projecting the state onto the true (physical) ground state. 
This, however,  can be done with impunity:  the projection only affects the ES eigenvalues at momenta with zero-energy eigenstates, while  the primary interest of our discussion is on the  topologically protected surface modes which live at momenta where the bulk spectrum is gapped. 
Further details on the projection and its effects on the ES can be found in Appendix \ref{app:projection}.

% - % - % - % - % - % - % - % - % - % - % - % - % - % - % - % - % - % - % - % - % - % - % - % - % - % - % - % - % - % - % - % - % - % - 

\section{Entanglement spectrum and surface states}\label{sec:ESvsSurface}
We now consider the properties of entanglement spectra for gapless three-dimensional KSLs, and how they are related to the surface spectra. 
We restrict our analysis to the ground state sector and focus on the ES of the resulting free Majorana system.  \footnote{Introducing other interactions, such as the Heisenberg spin exchange, generically mixes in fluxes into the ground state, as well as introduces interactions between the Majorana fermions. Both have interesting consequences for the spin liquid, see e.g. Ref.s~\cite{Song2016low,Hermanns2015spin-peierls}, whose effects on the entanglement spectrum are, however,  beyond the scope of this manuscript.}
The different example KSLs are chosen to be representatives for the different types of topological bandstructures that the Majorana fermions form, ranging from (topological) Fermi surfaces \cite{Hermanns2014quantum}, over nodal lines \cite{Mandal2009exactly} and nodal chains \cite{Yamada2017crystalline}, to Weyl \cite{Hermanns2015weyl} and Dirac points \cite{Yamada2017crystalline}. 
Details on the different lattice structures and the properties of the corresponding KSLs can be found in the appendices and Refs.~\cite{Obrien2016classification,Yamada2017crystalline}.

For all the systems we analyze, we find that the ES  mimics the surface spectrum of the Majorana system, consistent with previous studies \cite{Yao2010entanglement,Fidkowski2010entanglement}. 
In particular, topologically protected zero-energy surface states always have a counterpart at $\zeta=1/2$ in the ES. 
In fact, one can show the correspondence between surface and entanglement spectra by considering lower-dimensional gapped subsystems and using the results of Ref.~\cite{Fidkowski2010entanglement}. 
While there are many similarities between entanglement spectra and surface spectra, one should not view the former as a proper spectrum. 
In particular, `entanglement bands' generically exhibit discontinuities at momenta where the bulk gap vanishes.

The first example we consider is the Kitaev model on the (10,3)b (hyperhoneycomb) lattice (see App.~\ref{sec:103b} for details on the lattice) \cite{Mandal2009exactly}, which is believed to be relevant for the material $\beta$-Li$_2$IrO$_3$ \cite{Takayama2015hyperhoneycomb}.
The solution of the bare Kitaev Hamiltonian reveals a nodal line at the Fermi level \cite{Schaffer2015topological}, as shown in Fig.~\ref{fig:hhoney}(a). When considering open boundary conditions along one direction, the projection of the nodal line into the corresponding surface Brillouin zone (BZ) is filled with zero-energy states, also called drumhead states~\cite{Burkov2011topological}. 
Fig.~\ref{fig:hhoney}(b) shows both the zero-energy surface states and the $\zeta=1/2$ entanglement modes for the bare Kitaev model for a real space cut along the $\mathbf{a_1}$-direction. 
Clearly, the two look  identical. 
If  time-reversal symmetry is broken, i.e. $\kappa\neq0$ in Eq.~\eqref{eq:mag}, the nodal line in the BZ gaps out except at two points which are topologically protected Weyl points (WPs) \cite{Wan2011topological}. 
The drumhead surface state gaps out into a line --- a Fermi arc --- connecting the projections of the WP \cite{Hermanns2015weyl}. 
The same happens for the ES, c.f.\,Fig.~\ref{fig:hhoney}(c). 
Note that the Fermi arc is doubly degenerate in both cases. 
The close similarity between zero-energy surface modes and $\zeta=1/2$ entanglement modes is a generic feature that occurs in all systems that we analyzed. 
\begin{figure}[t]
	\centering
	\includegraphics{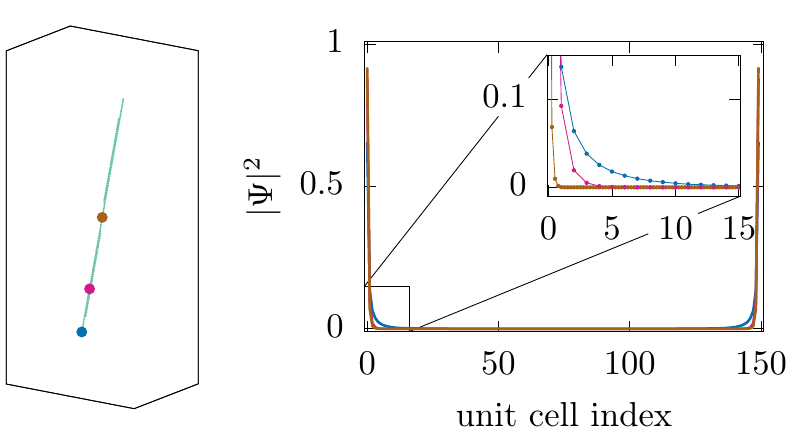}
	\caption{
		Localization of the entanglement eigenstates for three different momenta along the `entanglement Fermi arc', which are indicated in the left panel. The right panel shows the weight of the eigenstates as a function of the unit cell index, where the entanglement cuts are located at unit cells 0  and 150. All eigenstates show an exponential localization at the entanglement cuts with a localization length that depends on the distance to the WPs (see inset).
}
	\label{fig:prob}
\end{figure}

\begin{figure*}[ht]
	\centering
	\includegraphics{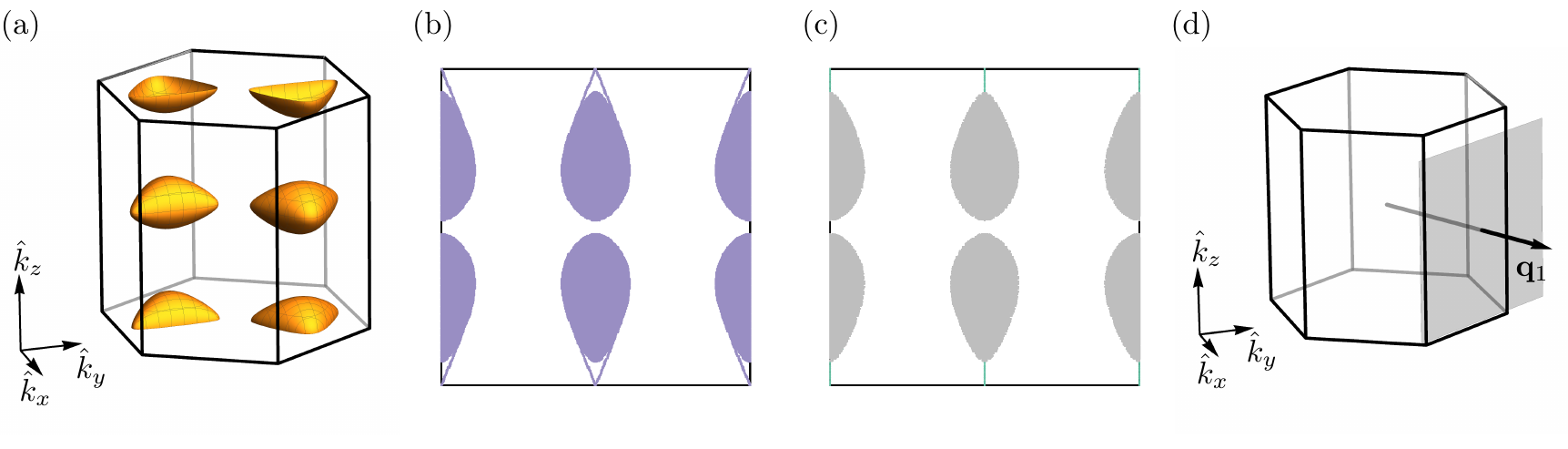}
	\caption{
		(a) Full BZ of the (8,3)a lattice. The Kitaev model with couplings $J_x=J_y=1.1$ and $J_z=0.8$ harbors four topological Fermi surfaces. 
		Panel (b) shows the zero-energy states of the surface spectrum and (c) the $\zeta=1/2$ states of the ES. In both cases, the Fermi arcs connecting the topological Fermi surfaces of opposite charge are clearly visible. 
		The surface BZ for open boundary conditions along $\mathbf a_1$ is shown in (d).  }
	\label{fig:83a}
\end{figure*}

Fermi arcs are located at the surface of the system, and the same holds for the entanglement Fermi arcs. 
We exemplify this feature using the  arcs of Fig.~\ref{fig:hhoney}(c). 
For different momenta along the arc, we plot the probability $
\lvert\Psi\rvert^2=\lvert\Psi_1\rvert^2+\lvert\Psi_2\rvert^2$ of the two states with entanglement eigenvalue $\zeta=1/2$ as a function of the position in the subsystem $A$. 
For all positions, the states are exponentially localized at the cut, as shown in the right panel of Fig.~\ref{fig:prob}.  
When approaching the projection of the WP at either end of the arc, the localization length diverges and the modes penetrate further and further into the bulk.

In the previous example, surface and entanglement spectra looked fully alike, but that is not a general feature. 
A simple counter-example is a topological metal --- realized for instance for the KSL on the (8,3)a lattice~\cite{Obrien2016classification} --- where each of the Fermi surfaces  surrounds a WP [see Fig.~\ref{fig:83a}(a)].  
Due to the WPs,  there are chiral surface states --- the remnants of the Fermi arcs --- that connect Fermi surfaces with opposite Weyl charge.
In the surface spectrum, the arcs are generically split because of the finite energy of the WPs [see  Fig.~\ref{fig:83a}(b)]. 
If one of the WPs sits at $E>0$, its particle-hole conjugate partner sits at $E<0$. This leads to the special feature that the Fermi arcs have to cross each other at some point in the BZ. For the KSL on the (8,3)a lattice, this occurs at the boundaries of the surface BZ. 
The ES, on the other hand, is oblivious to the energy of the WP, and the Fermi arcs still lie on top of each other [see Fig.~\ref{fig:83a}(c)].

Up until now, we only presented numerical evidence for the close relation between zero-energy surface modes and $\zeta=1/2$ entanglement modes, but one can in fact show it rigorously using the topological invariants that protect the gapless modes. 
To do so, we interpret the gapless three-dimensional system as a one- (or two-)dimensional one, where two (or one) of the momenta are regarded as parameters that tune between the gapped topological and trivial phase. 
The first harbors protected surface modes, the latter does not. 
For these gapped phases, one can then use the results by Fidkowski \cite{Fidkowski2010entanglement}, who found a one-to-one correspondence between the surface and ES.  

Let us make this more explicit by first considering a system with WPs, which are protected by the Chern number. 
We regard the three-dimensional system as a two-dimensional one, with one of the momenta being a tunable parameter. 
Depending on its value, the two-dimensional system is either in a trivial or a Chern insulator phase. 
The latter has an odd number of chiral edge modes that connect the valence and conduction band, while the former harbors an even number (most often 0) of such bands. 
Consequently, when we tune the parameter into the Chern insulator regime, there has to be a zero-energy mode at the surface~\cite{Wan2011topological}. 
Using the equivalency between the surface and ES for Chern insulators~\cite{Fidkowski2010entanglement}, we conclude that also the ES harbors an odd number of $\zeta=1/2$ modes when the momentum parameter is tuned to lie in the Chern insulator regime. 
The entanglement Fermi arcs terminate at the WPs, i.e. at the transition point between trivial and Chern insulator, in full analogy to the usual Fermi arcs. 

\begin{figure}[t]
	\centering
	\includegraphics[width=\columnwidth]{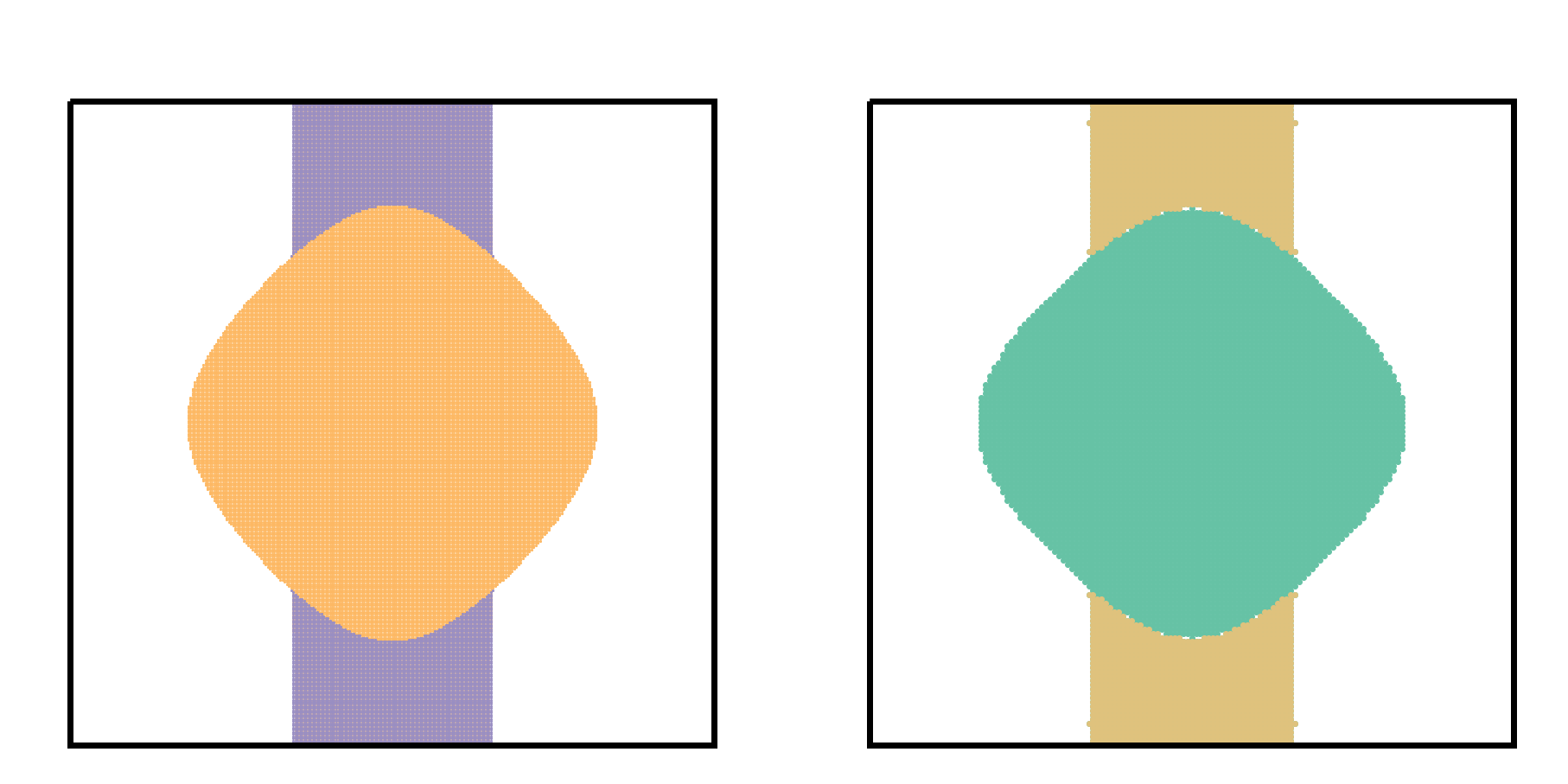}
	\caption{The zero-energy modes of the surface spectrum (left) and $\zeta=1/2$ modes of the ES (right) for a cut along $\mathbf a_3$-direction for the Kitaev model on the (10,3)d lattice. The modes marked in orange/green are two-fold, those marked in purple/yellow are four-fold degenerate.}
	\label{fig:103dsurface}
\end{figure}

\begin{figure*}[htb]
	\centering
	\includegraphics{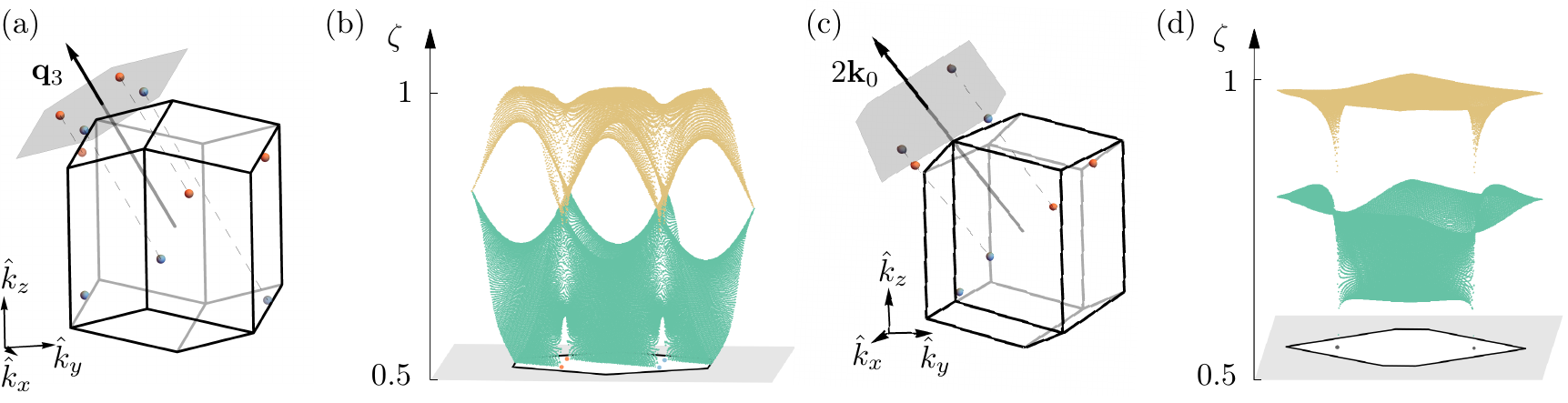}
	\caption{(a) Full BZ of the (8,3)b lattice for a real-space cut along $\mathbf{a}_3$;  the plot of the two lowest entanglement bands above $\zeta=1/2$ is shown in (b), with two Fermi arcs visible. 
		(c) Full BZ of the (8,3)b lattice for a real-space cut along $\mathbf a^*_1$.
		The plot of the four lowest bands above $\zeta=1/2$ is shown in (d). Note that each band is doubly degenerate. Even though there are no modes at $\zeta=1/2$, the hybridized Fermi arc is nevertheless clearly visible in the spectrum. 
	 }
	\label{fig:83ba1}
\end{figure*}

To explain the appearance of drumhead states, we first note that nodal lines are generically protected by time-reversal symmetry.  Given this symmetry, the Majorana Hamiltonian can be written in an off-block diagonal form as 
\begin{align}
H(\mathbf k)&=\left(      \begin{array}{cc} \mathbf 0 & \mathbf Q \\ \mathbf Q^\dagger & \mathbf 0    \end{array}  \right). 
\end{align}
We can now define the chiral invariant as 
\begin{align}\label{eq:chiralinv}
\theta=\frac{1}{4\pi i}\int_\ell dk \mbox{ tr}\left[\mathbf Q^{-1} \partial_k \mathbf Q-(\mathbf Q^{-1} \partial_k\mathbf Q)^\dagger \right],
\end{align}
where the integration is along a closed loop $\ell$ in the BZ, i.e. we compute the invariant along an effectively one-dimensional gapped system. 
The chiral invariant $\theta$ can take integer values. 
In the following, we focus on situations where the integration parameter $k$ lies on a non-contractible closed path, parallel to one of the reciprocal lattice vectors. 
Having a non-zero chiral invariant for, say, integrating along $k_3$ implies that there must be zero-energy surface modes as soon as the system is no longer periodic in the $\mathbf{a}_3$-direction.
 The number of edge modes is equal (or larger) than the bulk invariant. 
For the pure Kitaev model on the hyperhoneycomb lattice, the chiral invariant is $\pm 1$ if the integration path goes through the nodal line and 0 if it does not. Thus, the line is filled with (doubly-degenerate) zero-energy surface modes when the system is no longer periodic in the $\mathbf{a}_3$-direction. 
 Again utilizing the equivalence between surface and ES, we conclude that also the ES shows doubly degenerate  modes at $\zeta=1/2$ (located on opposite entanglement surfaces of the system) within the nodal line. 
A slightly more complex example of drumhead states is provided by the Kitaev model on the (10,3)d lattice, which harbors nodal chains. 
The chiral invariant in this system can either be 0, 1, or 2; the corresponding surface spectrum/ES degeneracy is 0, 2, and 4, respectively, as visualized in Fig.~\ref{fig:103dsurface}.

\section{Weak topological features in the entanglement spectrum}
\label{sec:topfeatures}
Interestingly, the ES can show features that are not visible in the surface spectrum. 
These are usually features that are not topological themselves, but have a topological origin. 
We exemplify this behavior by considering KSLs with Weyl and Dirac nodes, but it may also occur in other systems, such as gapped spin liquids or spin liquids with a Majorana Fermi surface. 

One simple way to identify Weyl spin liquids is by the presence of Fermi arcs in the surface spectrum. 
However, if one chooses the surface termination such that  WPs of opposite chirality are projected to the same momentum in the surface BZ, the resulting Fermi arcs are no longer topologically protected. 
For system where the corresponding Fermi arcs are not located at the same momenta, stable remnants of the Fermi arcs can still persist, even though they generically do not connect to the WPs any longer~\cite{Dwivedi2017}. 
However, additional symmetries can force the entire Fermi arcs to lie on top of each other, in which case they can hybridize and fully gap out. 
This implies that the resulting surface spectrum has usually no accessible information about the topological properties of the system. 
In the following, we show that the ES, on the other hand, still shows clear signatures of the hybridized Fermi arcs. 

Our first example is the Kitaev model on the (8,3)b lattice, which harbors a Weyl spin liquid even in the presence of both  time-reversal and inversion symmetry~\cite{Obrien2016classification}. 
Solving the Kitaev Hamiltonian for this lattice structure we find two pairs of WPs at the Fermi level, shown in Fig.~\ref{fig:83ba1}(a). 
Imposing open boundary conditions along the $\mathbf{a}_3$-direction the surface spectrum  shows Fermi arcs~\cite{Obrien2016classification}. 
The corresponding ES is plotted in Fig.~\ref{fig:83ba1}(b).
For reasons that will become clear later, we do not restrict our plot to the $\zeta=1/2$ modes as we did earlier, but instead plot the two lowest ES band that lie above $\zeta=1/2$. 
Higher bands are ignored as they all lie at $\zeta\approx 1$; the behavior of lower bands can be infered by using that for such half-filled systems, each entanglement energy at $\zeta$ implies one at $1-\zeta$. 
The band indicated in green shows the two entanglement Fermi arcs at $\zeta=1/2$. 
In this surface termination, the surface and ES contain the same topological information.  

\begin{figure}[t]
	\centering
	\includegraphics[width=\columnwidth]{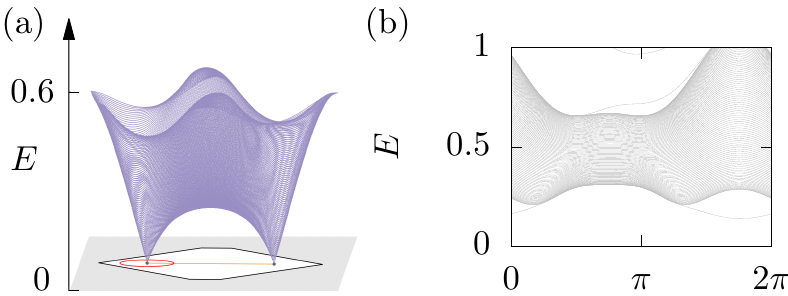}
	\caption{Surface spectrum of the Kitaev model on the (8,3)b lattice when cutting along $\mathbf{a}^*_1$. 
		(a) lowest energy band in the BZ, the orange line indicates the location of the hybridized Fermi arc. The red line indicates the momenta, for which the full spectrum is plotted in panel (b).  }
	\label{fig:83bsurface}
\end{figure}

We now choose a surface termination such that the WPs are projected onto the same point in the corresponding surface BZ [see Fig.~\ref{fig:83ba1}(c)]. 
For the (8,3)b lattice this is achieved by re-defining the real space lattice vectors (denoted with $^*$) as
\begin{align}
\mathbf{a}^*_1=\mathbf{a}_1,\quad\mathbf{a}^*_2=\mathbf{a}_2,\quad\mathbf{a}^*_3=\mathbf{a}_1-\mathbf{a}_3.
\end{align}
The original choice of the lattice vectors can be found in App.~\ref{app:83b}.
For open boundary conditions along the $\mathbf{a}^*_1$-direction the corresponding surface BZ is perpendicular to the vector $\mathbf{k}_0=\mathbf{q}_1/2-\mathbf{ q}_3/2$ which maps  WPs of opposite chirality onto each other. 
In the surface spectrum, the Fermi arcs hybridize and become masked by the bulk spectrum, as shown in Fig.~\ref{fig:83bsurface}.
In Fig.~\ref{fig:83bsurface}(a) we show the lowest energy band of the surface spectrum, with the expected position of the Fermi arc indicated in orange. 
We also plot the entire low-energy surface spectrum along the momenta indicated by the red ring in  Fig.~\ref{fig:83bsurface}(b).
The surface modes are still below the bulk modes,  but their hybridization gap is comparable to the band gap.
In fact, considering only the surface  spectrum in Fig.~\ref{fig:83bsurface}, one could never deduce that the system has a topological band structure. 

The ES for the same surface termination is shown in  Fig.~\ref{fig:83ba1}(d). 
Again, we plotted the two lowest bands above $\zeta=1/2$. 
For this surface termination, the green band does not contain modes at $\zeta=1/2$ any longer --- also the entanglement Fermi arcs have hybridized. 
In contrast to the surface spectrum however, the green band is also clearly separated from the bulk bands (i.e. the yellow band and all higher bands), 
and exists in an entanglement energy region that is usually reserved for topologically protected bands. 
It is, in fact, rather suggestive from Fig.~\ref{fig:83ba1} to interpret the green band (and its counterpart at $\zeta<1/2$ as hybridized entanglement Fermi arcs. 
Thus, from the ES we would still deduce that the system in question is special and has some topological properties.

\begin{figure}[t]
	\centering
	\includegraphics[width=.8\columnwidth]{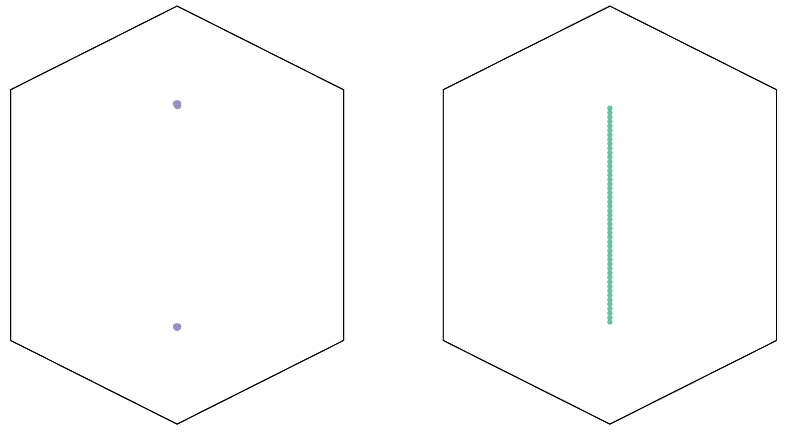}
	\caption{Zero-energy modes of the surface  (left) and $\zeta=1/2$ modes of the entanglement (right) spectrum of the Kitaev model on the \latname~lattice for a real-space cut along the $\mathbf{a}_1$-direction. }
	\label{fig:83Xsurface}
\end{figure}

Our second example concerns the Kitaev model on the \latname~lattice, where a four-fold screw symmetry stabilizes three-dimensional Dirac nodes~\cite{Yamada2017crystalline}. 
These can be thought of as a combination of pairs of oppositely charged WPs. 
Due to particle-hole symmetry in KSLs, the Fermi arc always connects WPs at $\mathbf{k}$ and $-\mathbf{k}$. 
One may wonder whether the surface/entanglement spectrum can distinguish this connectivity from the trivial one where the two WPs at $\mathbf{k}$  are connected to each other, and similarly the two at ($-\mathbf{k}$). 
When computing the surface spectrum for this system, we again find that the surface modes are masked by the bulk spectrum, and there is no clear evidence of the nontrivial Berry connection. 
In the ES, the  Fermi arcs hybridize, but the hybridization gap is extremely small and the modes basically remain at $\zeta=1/2$ (see Fig.~\ref{fig:83Xsurface}). 
One can therefore identify the nontrivial Berry connection very easily in the ES, while it is completely masked in the surface spectrum.

While one needs further study to determine for which types of topological features the ES is better suited than the surface spectrum, our examples  show that there are several situations where the ES contains more information about the topology of the system than the surface spectrum.

\section{Conclusion}
In this paper, we have computed the ES for a family of gapless KSLs, and  demonstrated that the entanglement spectrum contains valuable information even for gapless strongly correlated systems. 
Similar to  gapped systems, we find that the entanglement spectrum is closely related to the topologically protected surface spectrum. 
In most cases, the two spectra have a one-to-one correspondence, and contain exactly the same information about the topology of the state. 
Of the two, the ES is usually easier to interpret, for the following reason: a surface spectrum can accidentally have modes that are close to zero energy, but are not of topological origin. 
In the ES, on the other hand, there are no such accidental modes. 
In addition, in a surface spectrum, bulk bands may partially mask the dispersion of topological surface modes. 
In the ES,  the bulk modes always sit near $\zeta=0,1$  -- so that the  topological bands located close to $\zeta=1/2$  are always clearly visible and easy to identify.

In some cases, an even more interesting situation occurs: the ES can contain  more information than the corresponding surface spectrum. 
In particular, the ES is sensitive to hybridized topological surface states, such as the hybridized Fermi arcs discussed earlier. 
Such features are usually hidden in real surface spectra, because the hybridization gap can easily be larger than the bulk gap, so that the surface modes become completely masked by the bulk modes. 
In the ES, there are no bulk modes between $\zeta=0$ and $1$, so the hybridized modes are still easily distinguishable. 
This shows that the ES can be  a more powerful tool than the usual surface spectrum to identify and detect whether or not a system is topological.
Our results can easily be generalized to other noninteracting systems, in particular to topological semi-metals (see for instance Ref.~\cite{Chiu2016classification} and references therein). \\

\noindent {\it Acknowledgments.--}
We thank Simon Trebst for useful discussions.
The numerical simulations were performed on the CHEOPS cluster at RRZK Cologne.
S.M.\,was supported by the Deutsche Forschungsgemeinschaft under grant no.\,SFB 1238 and by the EPSRC under grant no.\,EP/L015110/1. M.H.\,was funded by the Deutsche Forschungsgemeinschaft via the Emmy Noether programme with grant no.\,HE 7267/1-1. Open data compliance: This is a theory paper and all results are
reproducible from the given formulas.

%%% Appendix %%%%%%%%%%%%%%%%%%%%%%%%%%%%%%%%%%%%%%%%%%%%%%

\appendix

\section{(8,3)a}\label{sec:83a}

The (8,3)a structure is a hexagonal lattice with six sites per unit cell. They are located at
\begin{alignat}{3}
    \mathbf{r}_1&=\left(\frac{1}{2},\frac{\sqrt{3}}{10},0\right),&&\mathbf{r}_2=\left(\frac{3}{5},\frac{\sqrt{3}}{5},\frac{2\sqrt{2}}{5}\right),\notag\\
    \mathbf{r}_3&=\left(\frac{1}{10},\frac{3\sqrt{3}}{10},\frac{\sqrt{2}}{5}\right),\quad&&\mathbf{r}_4=\left(\frac{2}{5},\frac{\sqrt{3}}{5},\frac{\sqrt{2}}{5}\right),\notag\\
    \mathbf{r}_5&=\left(0,\frac{2\sqrt{3}}{5},0\right),&&\mathbf{r}_6=\left(-\frac{1}{10},\frac{3\sqrt{3}}{10},\frac{2\sqrt{2}}{5}\right).
\end{alignat}
\begin{figure}[t]
	\centering
	\includegraphics[width=\columnwidth]{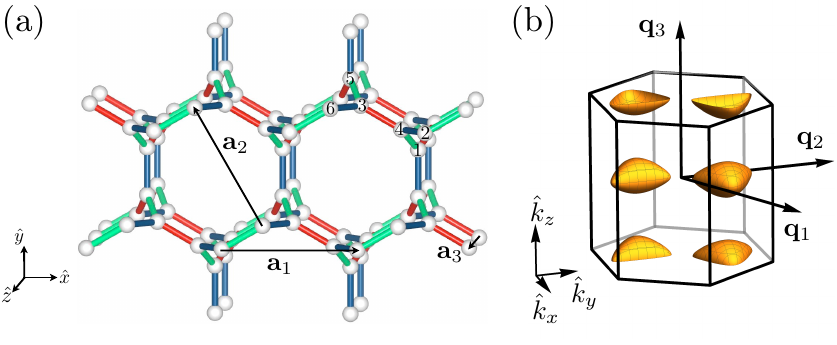}
	\caption{(a) Unit cell and translation vectors of the (8,3)a lattice. Green/red/blue bonds denote $x$-/$y$-/$z$-bonds. 
	(b) BZ with Majorana Fermi surfaces for Kitaev couplings $J_x=J_y=1.1$ and $J_z=0.8$. }
	\label{fig:BZ83a}
\end{figure}
The lattice translation vectors for the (8,3)a lattice are chosen as
\begin{align}
    \mathbf{a}_1&=\left(1,0,0\right),\quad\mathbf{a}_2=\left(-\frac{1}{2},\frac{\sqrt{3}}{2},0\right),\notag\\
    \mathbf{a}_3&=\left(0,0,\frac{3\sqrt{2}}{5}\right),
\end{align}
and the corresponding reciprocal lattice vectors are given by
\begin{align}
    \mathbf{q}_1&=\left(2\pi,\frac{2\pi}{\sqrt{3}},0\right),\quad\mathbf{q}_2=\left(0,\frac{4\pi}{\sqrt{3}},0\right),\notag\\
    \mathbf{q}_3&=\left(0,0,\frac{5\sqrt{2}\pi}{3}\right).
\end{align}
In Fig.~\ref{fig:BZ83a}(a) the unit cell of the lattice with the  lattice translation vectors is shown. Additionally, the bond colors indicate the Kitaev exchange interaction along a particular bond ($xx$ - green, $yy$ - red and $zz$ - blue). The bond operators $u_{jk}$ are defined as in~\cite{Obrien2016classification}, such that the system is in the ground state flux sector of the $\mathbb{Z}_2$ gauge field.
The solution of the bare Kitaev model on this lattice exhibits four Majorana Fermi surfaces in the gapless phase, shown in Fig.~\ref{fig:BZ83a}(b).
 For our calculation of the surface and ES we chose the following parameters: $J_z=0.8$, $J_x=J_y=(3-J_z)/2$. For this choice the system is within the gapless phase and exhibits stable Majorana Fermi surfaces. Within each surface sits one WP such that we can assign a topological charge to the Fermi surface. Changing the couplings constants leads to a deformation of the Fermi surfaces and in some parameter regions the Fermi surfaces are topologically trivial \cite{Obrien2016classification}. For the calculation of the ES and the surface spectrum in Fig.~\ref{fig:83a} we choose a real space cut along the $\mathbf{a}_1$-direction and the system size $L=200$ with an entanglement cut at $L/2$.

\section{(8,3)b}
\label{app:83b}
\begin{figure}[t]
	\centering
	\includegraphics[width=\columnwidth]{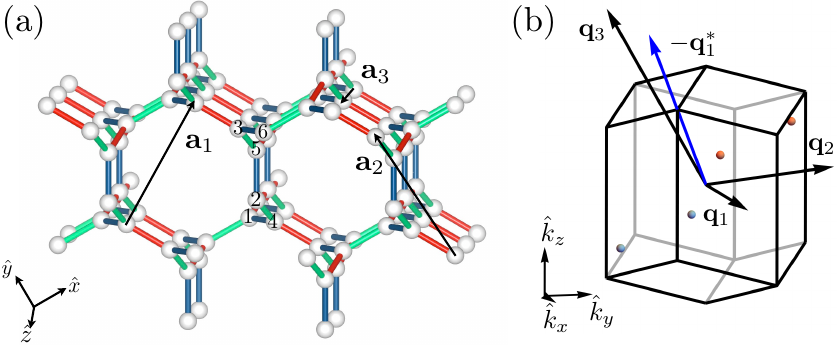}
	\caption{
		 (a) Unit cell and translation vectors of the (8,3)b lattice. Green/red/blue bonds denote $x$-/$y$-/$z$-bonds. 
		(b) BZ with the four WPs for  isotropic couplings $J_x=J_y=J_z=1$. }
	\label{fig:BZ83b}
\end{figure}
As the (8,3)a lattice the (8,3)b structure has a hexagonal geometry. In each unit cell there  six basis sites located at
\begin{alignat}{3}
    \mathbf{r}_1&=\left(\frac{1}{10},\frac{1}{2\sqrt{3}},\frac{1}{5}\sqrt{\frac{2}{3}}\right),&& \mathbf{r}_2=\left(\frac{1}{5},\frac{\sqrt{3}}{5},\frac{\sqrt{6}}{5}\right),\notag\\
    \mathbf{r}_3&=\left(\frac{3}{10},\frac{11}{10\sqrt{3}},\frac{4}{5}\sqrt{\frac{2}{3}}\right),\quad&&\mathbf{r}_4=\left(\frac{1}{5},\frac{2}{5\sqrt{3}},\frac{2}{5}\sqrt{\frac{2}{3}}\right),\notag\\
    \mathbf{r}_5&=\left(\frac{3}{10},\frac{3\sqrt{3}}{10},\frac{\sqrt{6}}{5}\right),&&\mathbf{r}_6=\left(\frac{2}{5},\frac{1}{\sqrt{3}},\sqrt{\frac{2}{3}}\right).
\end{alignat}
The lattice translation vectors for the (8,3)b lattice are chosen as
\begin{align}
    \mathbf{a}_1&=\left(\frac{1}{2},\frac{1}{2\sqrt{3}},\frac{1}{5}\sqrt{\frac{2}{3}}\right),\quad\mathbf{a}_2=\left(0,\frac{1}{\sqrt{3}},\frac{2}{5}\sqrt{\frac{2}{3}}\right),\notag\\
    \mathbf{a}_3&=\left(0,0,\frac{\sqrt{6}}{5}\right),
    \label{eq:83bvec}
\end{align}
and the corresponding reciprocal lattice vectors are given by    
\begin{align}
    \mathbf{q}_1&=\left(4\pi,0,0\right),\quad\mathbf{q}_2=\left(-2\pi,2\sqrt{3}\pi,0\right),\notag\\
    \mathbf{q}_3&=\left(0,-\frac{4\pi}{\sqrt{3}},5\sqrt{\frac{2}{3}}\pi\right).
\end{align}
Solving the bare Kitaev model on this inversion symmetric lattice we find four WPs at the Fermi level. 
The location of the gapless modes is shown within the BZ in Fig.~\ref{fig:BZ83b}(b),  next to the lattice with the unit cell and translation vectors shown in (a). 
Within the BZ the vector $\mathbf{k}_0=\mathbf{q}_1/2-\mathbf{q}_3/2$ maps a pair of WPs onto each other. 
For a different choice of lattice vectors
\begin{align}
\mathbf{a}_1^*=\mathbf{a}_1,\quad\mathbf{a}_2^*, \quad\mathbf{a}_3^*=\mathbf{a}_1-\mathbf{a}_3,
\label{eq:83balt}
\end{align}
 the corresponding reciprocal lattice vectors become
\begin{align}
\mathbf{q}_1^*=\left(-4\pi,\frac{4\pi}{\sqrt{3}},-5\pi\sqrt{\frac{2}{3}}\right),\quad\mathbf{q}_2^*=-\mathbf{q}_2,\quad\mathbf{q}_3^*=\mathbf{q}_3.
\end{align}
Thus, a cut along the $\mathbf{a}_1^*$-direction results in a surface BZ that is perpendicular to the vector $\mathbf{k}_0$. 
Within the surface BZ WPs of opposite chiralites are mapped onto each other, leading to a gap in the ES, c.f.\,Fig.~\ref{fig:83ba1}.
Our numerical simulations were all performed within the gapless phase at the isotropic point of the exchange couplings. 
The system size for the spectra for a cut along $\mathbf{a}_3$-direction ($\mathbf{a}_1^*$) is $L=150$ ($L=198$).
The larger system size for the $\mathbf{a}_1^*$-direction was necessary to resolve the discontinuity of the spectrum around the projections of the WPs.
\section{$8^2.10$-$a$}

\begin{figure}[t]
	\centering
	\includegraphics[width=\columnwidth]{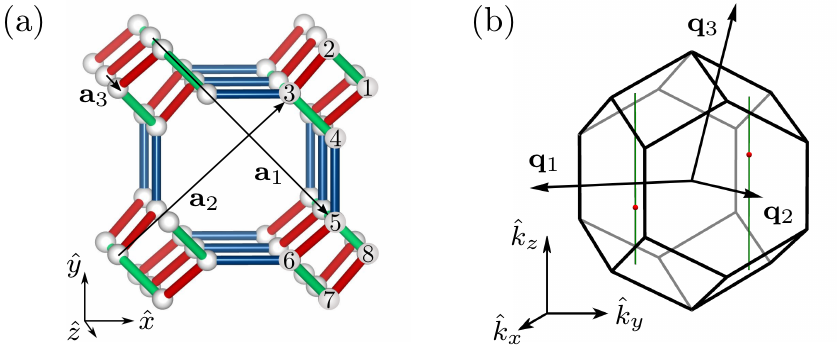}
	\caption{(a) Unit cell and translation vectors for the ~$8^2.10$~-~$a$ lattice. Green/red/blue bonds denote $x$-/$y$-/$z$-bonds. 
		(b) BZ of the $8^2.10$~-$a$ lattice with the two gapless Dirac nodes indicated by red points. The two green lines denote the momenta that are invariant under the four-fold screw symmetry that protects the Dirac nodes. }
	\label{fig:BZ8x}
\end{figure}
The $8^2.10$-$a$ lattice is inversion symmetric, and invariant under a four-fold screw symmetry when $J_x=J_y$. 
The eight sites of the unit cell are located at
\begin{alignat}{3}
    \mathbf{r}_1&=\frac{1}{4}(a,b,1),&&\mathbf{r}_2=\frac{1}{4}(0,a+b,2),\notag\\
    \mathbf{r}_3&=\frac{1}{4}(-a,b,3),&&\mathbf{r}_4=\frac{1}{4}(0,-a+b,4),\notag\\
    \mathbf{r}_5&=\frac{1}{4}(0,a-b,4),&&\mathbf{r}_6=\frac{1}{4}(-a,-b,3),\notag\\
    \mathbf{r}_7&=\frac{1}{4}(0,-a-b,2),\quad&&\mathbf{r}_8=\frac{1}{4}(a,-b,1),
\end{alignat}
with $a=1/\sqrt{2}$ and $b=2a$. 
The lattice translation vectors are
\begin{align}
    \mathbf{a}_1&=\frac{1}{2}(b,-b,1),\quad\mathbf{a}_2=\frac{1}{2}(b,b,1),\notag\\
    \mathbf{a}_3&=(0,0,1),
\end{align}
and the corresponding reciprocal vectors are given by
\begin{align}
    \mathbf{q}_1&=\left(\frac{2\pi}{b},-\frac{2\pi}{b},0\right),\quad\mathbf{q}_2=\left(\frac{2\pi}{b},\frac{2\pi}{b},0\right),\notag\\
    \mathbf{q}_3&=\left(-\frac{2\pi}{b},0,2\pi\right).
\end{align}

The Kitaev model is unusual in that the Majorana system hosts \emph{Dirac} nodes as long as the four-fold screw rotation and time-reversal symmetry are not broken~\cite{Yamada2017crystalline}. The lattice, with unit cell and translation vectors is shown in Fig.~\ref{fig:BZ8x}(a). 
The locations of the Dirac nodes at  the isotropic point $J_x=J_y=J_z=1$ are visualized in Fig.~\ref{fig:BZ8x}(b). The plots in the main text are also done for the isotropic couplings for a system size of $L=150$.

\section{(10,3)b}\label{sec:103b}
\begin{figure}[t]
\centering
\includegraphics[width=\columnwidth]{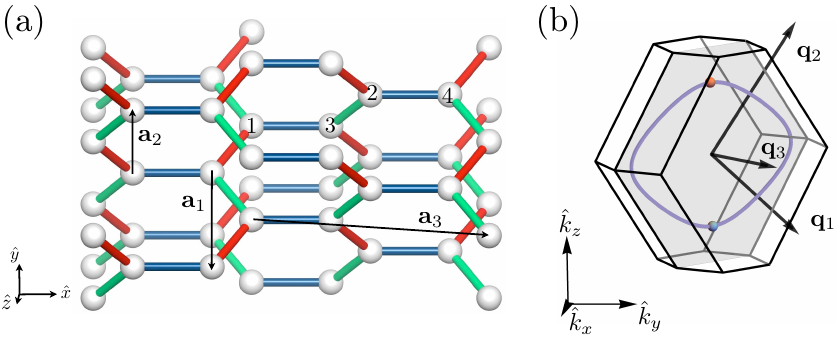}
\caption{(a) Unit cell and translation vectors for the Kitaev model on the  (10,3)b (hyperhoneycomb) lattice.
	Green/red/blue bonds denote $x$-/$y$-/$z$-bonds. 
	 (b) The BZ with the nodal line indicated in purple for isotropic couplings $J_x=J_y=J_z$. When time-reversal symmetry is broken, the nodal line gaps out except for two  WP (marked in orange/blue) at the Fermi level.}
\label{fig:BZ103b}
\end{figure}
The (10,3)b lattice, also referred to as hyperhoneycomb lattice, is a tetragonal lattice with four sites within each unit cell at
\begin{align}
    \mathbf{r}_1&=(0,0,0),\quad\mathbf{r}_2=(1,2,1),\notag\\
    \mathbf{r}_3&=(1,1,0),\quad\mathbf{r}_4=(2,3,1).
\end{align}
We choose the lattice translation vectors as
\begin{align}
\mathbf{a}_1=(-1,1,-2),\quad\mathbf{a}_2=(-1,1,2),\quad\mathbf{a}_3=(2,4,0),
\end{align}
and the corresponding reciprocal lattice vectors are
\begin{align}
    \mathbf{q}_1&=\left(-\frac{2\pi}{3},\frac{\pi}{3},-\frac{\pi}{2}\right),\quad\mathbf{q}_2=\left(-\frac{2\pi}{3},\frac{\pi}{3},\frac{\pi}{2}\right),\notag\\
    \mathbf{q}_3&=\left(\frac{\pi}{3},\frac{\pi}{3},0\right).
\end{align}
In Fig.~\ref{fig:BZ103b}(a) the four-site unit cell is shown with colored bonds representing the exchange between neighbors.
The solution of the bare Kitaev Hamiltonian reveals a nodal line at the Fermi level>\cite{Schaffer2015topological}. At the isotropic point $~J_x=J_y=J_z=1$ the nodal lines lies in the $k_x+k_y=0$ plane within the BZ, indicated by the gray shaded plane in Fig.~\ref{fig:BZ103b}(b). \\
If the time-reversal symmetry breaking term \eqref{eq:mag} is included in the model, i.e. $\kappa\neq0$, the nodal line gaps out except at two points which are topologically protected WPs at the Fermi level \cite{Hermanns2015weyl}. In our numerical calculation we chose $\kappa=0.1$.

The spectra in Fig.~\ref{fig:hhoney} were calculated for a system with length $L=200$ along the $\mathbf{a}_1$-direction at the isotropic point with an entanglement cut at $L/2$. 
\section{(10,3)d}
\begin{figure}[t]
	\centering
	\includegraphics[width=\columnwidth]{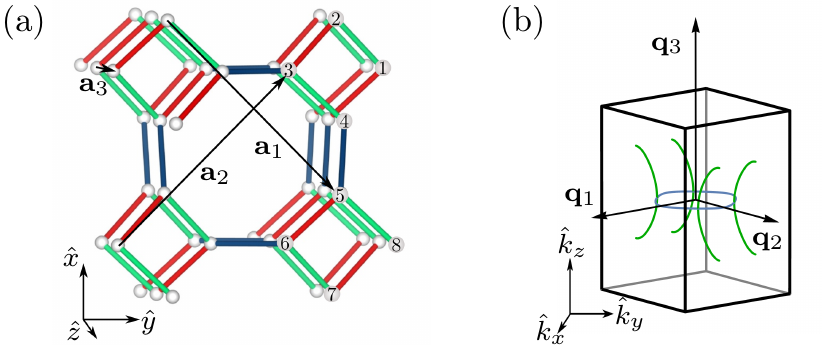}
	\caption{(a) Unit cell and translation vectors for the Kitaev model on the  (10,3)d lattice.
		Green/red/blue bonds denote $x$-/$y$-/$z$-bonds. 
		(b) The BZ with the three nodal lines indicated in blue/green for isotropic couplings $J_x=J_y=J_z$. When time-reversal symmetry is broken, the green nodal lines gap out, but the blue one remains stable.}
	\label{fig:BZ103d}
\end{figure}

The (10,3)d lattice is a primitive orthorombic structure with an eight-site unit cell. The site are located at
\begin{alignat}{3}
    \mathbf{r}_1&=\frac{1}{4}(a,b,1),&&\mathbf{r}_2=\frac{1}{4}(0,a+b,2),\notag\\
    \mathbf{r}_3&=\frac{1}{4}(-a,b,3),&&\mathbf{r}_4=\frac{1}{4}(0,-a+b,4),\notag\\
    \mathbf{r}_5&=\frac{1}{4}(0,a-b,3),&&\mathbf{r}_6=\frac{1}{4}(-a,-b,2),\notag\\
    \mathbf{r}_7&=\frac{1}{4}(0,-a-b,1),\quad&&\mathbf{r}_8=\frac{1}{4}(a,-b,4),
\end{alignat}
where $a=4-2\sqrt{2}$ and $b=2$. The  lattice translation vectors are
\begin{align}
    \mathbf{a}_1&=\frac{1}{2}(b,-b,0),\quad\mathbf{a}_2=\frac{1}{2}(b,b,0),\notag\\
    \mathbf{a}_3&=(0,0,1),
\end{align}

and the reciprocal lattice vectors are given by
\begin{align}
    \mathbf{q}_1&=\left(\frac{2\pi}{b},-\frac{2\pi}{b},0\right),\quad\mathbf{q}_2=\left(\frac{2\pi}{b},\frac{2\pi}{b},0\right),\notag\\
    \mathbf{q}_3&=(0,0,2\pi).
\end{align}
The lattice with the unit cell and translation vectors is shown in Fig.~\ref{fig:BZ103d}(a)
At the isotropic point, $J_x=J_y=J_z=1$, the solution of the bare Kitaev model exhibits three nodal lines that are linked, see Fig.~\ref{fig:BZ103d}(b).
The touching points are protected by the combination of time-reversal symmetry and a glide symmetry. The latter involves a reflection with respect to the $k_z=0$ plane and a subsequent translation within this plane~\cite{Yamada2017crystalline}. 
When considering open boundary conditions along $k_3$, one finds that there are three regions that are distinguished by their chiral invariant, as is visualized in Fig.~\ref{fig:103dsurface}. 
We studied the surface and ES for this geometry for a real-space cut along the $\mathbf{a}_3$-direction. The surface spectrum reveals gapless states filling the projections of the nodal lines, similar to the case of the (10,3)b (hyperhoneycomb) lattice. 
In our numercial simulation the spectra were calculated on systems of size $L=150$.

\section{Projection}\label{app:projection}

For all calculations in this manuscript, we used the unprojected ground state wave function. 
This is commonly done within the community of Kitaev spin liquids.  
Here, we want to comment on how projection affects the entanglement spectra, and in particular, which features are bound to be identical for the projected and the unprojected ground state. 

As shown in Ref.~\cite{Pedrocchi2011physical}, projecting the ground state from the extended Hilbert space to the physical subset is equivalent to enforcing a certain parity --- even or odd depending on the system size --- on the ground state.  
For the gapless systems, as  discussed here, it implies that the true ground state of the system is an equal weight superposition of all states with a given parity of physical, fermionic zero modes. 
In contrast, the `ground state' used in the manuscript at hand is a single state with unspecified parity. 
It is important to note that the effect of the projection is limited to momenta with zero-energy states, i.e. those momenta that constitute the Fermi surface, nodal lines, or Weyl/Dirac points.  
There is no disparity between the projected and unprojected state at momenta where the bulk spectrum is gapped: negative energy fermionic modes are filled, while positive ones remain empty. 
This readily implies that the two states --- projected and unprojected --- are identical at these momenta, though they may differ  when considering momenta with zero-energy modes. 

The main focus of the analysis of entanglement spectra lies in the analysis of virtual surface modes of gapless topological systems and their relation to  real surface modes. 
Those modes necessarily live at momenta where the bulk spectrum is gapped. 
Thus, the virtual surface modes remain unaffected by the projection to the physical subspace, and the main results and interpretations of our manuscript hold equally well for the true physical ground state of Kitaev spin liquids as for the unprojected ground state that we used.

%%% Bibliography %%%%%%%%%%%%%%%%%%%%%%%%%%%%%%%%%%%%%%%%%% 

\bibliographystyle{mybib}
\bibliography{references}

\end{document}